\documentclass[acmsmall,nonacm]{acmart}
\usepackage{fdiaz}
\usepackage{subcaption}
\usepackage{siunitx}
\usepackage[ruled]{algorithm2e} 
\usepackage{color,soul}

\AtBeginDocument{%
  }

\copyrightyear{2022}
\acmYear{2022}

\acmConference[]{}
\acmBooktitle{}

\newcommand{\ranking}[0]{\pi}
\newcommand{\rankings}[0]{\ranking}
\newcommand{\numusers}[0]{m}
\newcommand{\numitems}[0]{n}
\newcommand\recall{\operatorname{R}}
\newcommand{\user}[0]{u}
\newcommand{\users}[0]{\mathcal{U}}
\newcommand{\commonality}[0]{\text{C}}
\newcommand{\familiarity}[0]{F}

\newcommand{\catalog}[0]{\mathcal{D}}
\newcommand{\categories}[0]{\mathcal{G}}
\newcommand{\category}[0]{g}
\newcommand{\systems}[0]{\mathcal{S}}

\DeclareMathOperator{\prob}{Pr}

\newcommand{\ndcg}[0]{\mathrm{NDCG}}

\newcommand\andcg{\operatorname{\alpha-NDCG}}
\newcommand\iaerr{\operatorname{IA-ERR}}
\newcommand\precision{\operatorname{P}}
\newcommand{\eild}[0]{\mathrm{EILD}}
\newcommand{\epd}[0]{\mathrm{EPD}}

\newcommand{\dispexp}[0]{\mathrm{U}}

\newcommand{\dexpsq}[0]{\Delta_{\mathrm{sq}}}
\newcommand{\dexpabs}[0]{\Delta_{\mathrm{abs}}}
\newcommand{\dexpkl}[0]{\Delta_{\mathrm{KL}}}

\newcommand{\rel}[0]{\mathcal{R}}
\newcommand\ind{\operatorname{I}}
\newcommand\kld{\operatorname{D_{KL}}}

\DeclareMathOperator{\E}{\mathbb{E}}

\newcommand\rankeq{\mathrel{\stackrel{\makebox[0pt]{\mbox{\normalfont\tiny rank}}}{=}}}

\begin{document}
\title[Commonality in Recommender Systems]{Commonality in Recommender Systems: Evaluating Recommender Systems to Enhance Cultural Citizenship}

\author{Andres Ferraro}
\email{andresferraro@acm.org}
\orcid{0000-0003-1236-2503}
\affiliation{%
  \institution{McGill University}
  \city{Montr\'eal}
  \country{Canada}
}

\author{Gustavo Ferreira}
\email{gustavo.ferreira@mila.quebec}
\affiliation{%
  \institution{McGill University}
  \city{Montr\'eal}
  \country{Canada}
}

\author{Fernando Diaz}
\affiliation{
  \institution{Google}
  \country{Canada}
}
\email{diazf@acm.org}

\author{Georgina Born}
\affiliation{%
  \institution{University College London}
  \city{London}
  \country{United Kingdom}}
\email{g.born@ucl.ac.uk }

\begin{abstract}
Recommender systems have become the dominant means of curating cultural content, significantly influencing the nature of individual cultural experience. While the majority of academic and industrial research on recommender systems optimizes for personalized user experience, this paradigm does not capture the ways that recommender systems impact cultural experience in the aggregate, across populations of users. Although existing novelty, diversity, and fairness studies probe how recommender systems relate to the broader social role of cultural content, they do not adequately center culture as a core concept and challenge. In this work, we introduce commonality as a new measure of recommender systems that reflects the degree to which recommendations familiarize a given user population with specified categories of cultural content. Our proposed commonality metric responds to a set of arguments developed through an interdisciplinary dialogue between researchers in computer science and the social sciences and humanities. With reference to principles underpinning public service media (PSM) systems in democratic societies, we identify universality of address and content diversity in the service of strengthening cultural citizenship as particularly relevant goals for recommender systems delivering cultural content. We develop commonality  as a measure of recommender system alignment with the promotion of content toward a shared cultural experience across a population of users.  We advocate for the involvement of human editors accountable to a larger value community as a fundamental part of defining categories in the service of cultural citizenship.  As such, we ensure that commonality emphasizes a \textit{diverse} shared experience.  We empirically compare the performance of recommendation algorithms using commonality with existing utility, diversity, novelty, and fairness metrics using three different domains. Our results demonstrate that commonality captures a property of system behavior complementary to existing metrics and suggests the need for alternative, non-personalized interventions in recommender systems oriented to strengthening cultural citizenship across populations of users.  Moreover, commonality demonstrates both  consistent results under different editorial policies and robustness to missing labels and users.  Alongside existing fairness and diversity metrics,  commonality contributes to a growing body of scholarship developing `public good' rationales for digital media and machine learning systems.

\end{abstract}



\maketitle
\section{Introduction}

Online platforms that host cultural content such as music, movies, and literature use recommender systems to suggest and distribute items from their catalogs using the principle of personalization.  Generally, we measure the degree to which a recommender system succeeds in personalization by adopting various offline metrics (e.g. precision, NDCG) and online metrics (e.g. clickthrough rate, consumption) \cite{chandar2020}. Evaluation using these metrics is appealing in commercial settings because they are aligned with revenue-generating metrics like retention and subscriptions. As a result, personalization remains a central principle of academic and industrial research on recommender systems.

However, increasing evidence suggests that, while the degree of personalization is one desirable property of a recommender system, it does not capture the wider effects of recommender systems in aggregate, nor does it measure the effects of recommender systems across a population of users.  This is important because personalized recommendations are likely to have cumulative effects, shaping the wider cultures and societies within which they are being used \cite{anderson2020}.

We advocate that the design of recommender systems delivering cultural content broaden its foundation to include not just  personalization and associated commercial interests but also appropriate normative principles oriented to furthering the democratic well-being and the cultural and social development of contemporary societies. By `normative' we refer to principles considered to provide models of morally, ethically and/or politically right or just action or behavior at the level of societies and communities as well as individuals. And, just as domains such as criminal justice or lending have associated normative values related to justice and fairness, the distribution of cultural content does as well.  For guidance on normative principles appropriate to the provision of cultural content we turn to the principles underpinning public service media (PSM) systems~\cite{andrejevic2013public}. Public service media refers to the existence of various channels of content distribution and related media organizations that are designed to be accountable to the public and may also be publicly funded \cite{hendy:psb}.  Examples include the British Broadcasting Corporation, the Canadian Broadcasting Corporation, and the Australian Broadcasting Corporation.\footnote{PSM organizations are found throughout Europe and in many member states of the Commonwealth. Several PSM organizations may be found within a single country and thereby constitute a PSM ecology or system, as is the case, for example, in the UK, Germany, and Australia. Transnational PSM institutions also exist, such as the European Broadcasting Union, an alliance composed of PSM organizations from countries that lie within the European Broadcasting Area or are members of the Council of Europe.}

From the set of normative principles guiding PSM, we identify universality of address and content diversity in the service of strengthening cultural citizenship as particularly relevant normative principles for recommender systems delivering cultural content. If personalization attempts to maximize individual user satisfaction with a platform, the promotion of cultural citizenship entails disseminating a diversity or plurality of cultural content in order to stimulate intercultural and intracultural dialogue and exposure to cultural diversity.  As a result, the distribution of cultural content can enhance both social integration and pluralistic cultural experience across communities. In making these arguments we contribute to a growing body of scholarship developing public good rationales for digital media and machine learning systems~\cite{murdock2005building,andrejevic2013public,van2018public,moe2008dissemination,born:21stcentury,born:ecology,unterberger2021public}.  Later, we expand upon and further justify these arguments.

Recognizing the role played by evaluation metrics in embodying values such as personalization or fairness, we derive a new evaluation metric based on the principle of commonality.  Our metric measures the degree to which a recommender system familiarizes a given population of users with specific under-represented categories of cultural content in a certain medium. These categories, identified by human editors answerable to a knowledgeable community, work in concert with the metric to promote cultural citizenship.  More concretely, our commonality metric provides editors with an instrument to counteract undesirable biases associated with racism, sexism, and the neglect of non-Western content in the cultural content being recommended, and to deliver this more diverse experience commonly across a population of users.  

In order to better understand our metric, we include a series of quantitative analyses of its behavior.  Using data from three media---movies, music, and literature---we compare commonality with existing utility, diversity, and fairness metrics.  Our results demonstrate that our new metric is not correlated with existing metrics (i.e. it captures different properties) while maintaining comparable robustness.

To date, criticisms of recommender systems and machine learning systems for their capacity to reproduce forms of bias and discrimination have been based on the evaluation of such biases at the level of individual users. Relatively little attention has been paid to identifying means of both counteracting biases and enhancing diversity in recommended cultural content by evaluating their performance in the promotion of common experiences across a population of users. As we show in this work, recommender systems can be developed explicitly to promote a value such as diversity by counteracting racist and sexist biases and the neglect of non-Western content--and they can advance these progressive changes as common experiences, thus enhancing cultural citizenship. In this way recommender design, and evaluation in particular, can support the wider cultural changes called for by those critical of the lack of diversity and biases evident in recommender systems , as well as by those sympathetic to these criticisms from the RecSys community ~\cite{west2020redistribution, Benjamin2019, Noble2018, mansoury2020feedback, ekstrand2018, epps2020artist}.
\section{Background}
\label{sec:background}
Our research proceeded through sustained interdisciplinary dialogues between two computer science researchers designing music recommender systems (Diaz, Ferraro) and two media and communication scholars from the humanities and social sciences (Born, Ferreira). Over the course of a year we instructed each other in appropriate background research, sharing ideas and deepening our mutual engagement in both directions. In this way we translated terms from one `side' to the other, while also responding to critical questioning about the relevance of key concepts, and subsequently adapting the latter. Such translation across disciplinary domains is difficult and may be incomplete. Nonetheless, our experience is that it can produce hybrid thinking that can in turn generate powerful new concepts and tools. Indeed, systematic interdisciplinary practices of this kind can move beyond the tendency for one domain to provide merely a service to the other \cite{barryborn:interdisciplinarity}, and instead makes possible reflexive critical thought on both `sides' that builds towards new, higher-level syntheses.

\section{Four Propositions for Recommender Systems}
\label{sec:propositions}
The process described in Section \ref{sec:background} resulted in four related propositions at the core of our research, as follows.

First, we propose that it is timely for the design of recommender systems delivering cultural content to move beyond a commercial orientation focused primarily on individualized interests.\footnote{Although concerns about personalization often center on filter bubbles that keep individuals in taste echo chambers, our proposition in this paper goes further. We are concerned not only with the social effects of individualization but with the need for greater diversity in the curation and distribution of cultural content and the benefits of promoting a shared diversity of cultural experience.} We suggest that recommender design should, in addition, pursue complementary design paradigms guided by normative principles intended to promote the democratic development of contemporary cultures and societies as this enhances human flourishing. 

In this way, we link our work to `a computational politics wedded to emancipation and human flourishing'~\cite{stark2018algorithmic}.

Second, we propose that as well as a focus on personalization, recommender system design should acknowledge the aggregate and cumulative influences of recommender systems we have described, which have the potential to mediate wider cultural and social changes, and in this light develop ways of analyzing and modifying these influences in progressive ways that seek to achieve the goals described in the previous paragraph. In this respect our work participates in the `values in design' debate~\cite{Flanagan2008,knobel2011values}, which addresses the challenges of reflexively `incorporating human values adequately into formal models'~\cite{fish2021}. `Values in design' recognizes that the development of formal models for machine learning systems tend to fall back on `internalist' tendencies, in that `only considerations that are legible within the language of algorithms', for example accuracy and efficiency, `are recognized as important design and evaluation considerations'~\cite{green2020algorithmic}. The result is that design responses `to questions concerning human values such as fairness [become] problematic, because “problems with quantification [affect] everything downstream”'~\cite{fish2021,malik2020hierarchy}. It is in order to suggest a new approach to identifying values for recommender design that we turn below to studies of the principles guiding public service media systems. As ~\citet{fish2021} comment, `expanding formal models to include social values', what~\citet{green2020algorithmic} call “formalist incorporation”, may be `situationally and strategically useful', even if it is imperative to be aware of how `such solutions are insufficient as full remedies to the inherent limitations of formal modeling'~\cite{fish2021}.

Third, as a concrete means of addressing these two propositions, we turn to evaluation metrics as a place to incorporate alternative `values in design' into the development of recommender systems.  Specifically, we have developed a metric named `commonality' which measures the degree to which recommendations familiarize a given user population with specified categories of content chosen to promote certain `values in design' ~\cite{ferraro2022}. In addition to translating normative principles from the social sciences into mathematical representations, we conducted a series of experiments to assess the novelty and usefulness of commonality, in the context of other, related evaluation metrics.

Fourth, 

we draw on research that identifies the normative principles underlying public service media (PSM) systems, principles that then can be translated into a quantitative metric. In turning to previous research on principles embodied in PSM systems, we note both the powerful insights that can be derived from this research and its limitations with respect to the challenge of translating earlier normative concepts into contemporary digital platforms. Hence, although there is a literature concerned with how PSM organizations and older private media organizations are adapting to platformization and personalization (e.g., \cite{van2018public,hermida2010wikifying,candel2012adapting,bonini2017participatory,van2008can,syvertsen2019media}), as well as papers by PSM-based researchers on these topics and specifically on recommender design (e.g., \cite{fields2018,boididou2021building}, attempts to adapt PSM's earlier normative principles to the platform present are less advanced. We stress that, in this study, we are not concerned with PSM organizations in themselves nor with their approaches to recommender systems (see, e.g., \cite{jones:psm-recs}). Rather, we take writings on the normative foundations of PSM as a source of potentially relevant concepts, and then attempt to translate these relevant principles into the design of recommender systems. Our work does not aim to be an intervention in PSM but to have general implications for recommender design.

\section{The Case for Principles to Inform the Design of Recommender Systems}\label{sec:motivation}

Recommender systems have become the dominant means of curating cultural content in the digital era. Curation---or the selection, organization and promotion of content to be made available to consumers---has, however, a deep history. For centuries consumers have encountered cultural content through intermediaries---publishers, gallerists, patrons, impresarios---who collected works of culture, music and art, organized and categorized those works, and made them available to audiences and consumers. From the 2000s, the term curation began to be used to refer to the collection and organization of content on the internet. Indeed, the present has been depicted as an era of `curationism'---an `acceleration of the curatorial impulse to become a dominant way of thinking and being… [in] an attempt to make affiliations with, and to court, various audiences and consumers'~\cite{balzer2014curationism}. In general, `the introduction of new technologies has both introduced new methods of curation and expanded the breadth of individuals deemed fit to be curators'~\cite{alation:webpage}. Yet curation is not just an individual activity: it forges interconnections between curators, artists, and the industries, institutions, and online platforms supporting cultural production~\cite{obrist2008brief}. Today, the normative question of how curation by online platforms should be organized, and the forms it should take, is a pressing one.

When the curation enacted by online platforms' recommender systems is multiplied across the billions of recommendations presented to users, it significantly influences the nature of individual cultural experiences \cite{tomlein2021audit}. Yet in marked contrast with earlier eras of curation, this influence is multiplied and magnified cumulatively not only across time but across populations, cultures, and regions. In the short term, recommender systems clearly influence individual cultural consumption and taste. In the medium and long term, by employing data on consumer behavior and repeatedly influencing consumer choices, recommender systems can shape cultural literacies as well as population-wide trends in cultural consumption and cultural taste \cite{born:cifar}. They also participate in the commodification of the data generated by consumers as they engage with online platforms~\cite{fuchs2012}. Moreover, the collection and mining of consumer data implemented by recommender systems, a type of `monitoring-based marketing'~\cite{andrejevic2013public}, takes place at a much larger scale and is more rapid, recursive and intensive in comparison with earlier, non-computational methods of applying market research to identify and shape what consumers might want. In some ways this may be a productive kind of power and control exercised over consumers; yet, as a critic notes, `users have little choice over whether this data is generated and little say in how it is used', and `in this sense we might describe the generation and use of this data as… [an] alienated or estranged dimension of [users'] activity'~\cite{andrejevic2013public}.

Recommender systems therefore implement a higher degree of automatized intervention than previous forms of curation in the way not only individuals but societies and communities encounter cultural content. And despite their intended personalized address, recommender systems have cumulative effects in shaping the wider cultures and societies within which they are being employed. 
Yet perhaps because academic and industrial research on recommender systems has converged on personalization as a paradigm, these effects have been relatively unexplored by research in the RecSys community. 
Some metrics linked to personalization and `user relevance' (e.g., nDCG, precision, clickthrough rate) align with user retention and other longer term individual-level metrics which, when aggregated, can correlate with business metrics like revenue \cite{theocharous:ltv}.  Even metrics like diversity, often motivated by broadening the range of categories to which users are exposed, regularly need to be justified by individual-level metrics like retention \cite{anderson2020}.  Despite being concerned with and sensitive to the broader social role of cultural content, fairness metrics focus on individual exposure of providers or effectiveness for users and do not capture the wider, aggregate shaping effects of recommender systems on patterns of cultural consumption, taste, and literacy as described above.

\subsection{Applying normative principles from public service media to recommendation: universality, diversity, and cultural citizenship}
In employing the principles underpinning public service media systems in democratic societies, we take the view that `a public service rationale is as pertinent as ever in the digital era'~\cite{andrejevic2013public}. The normative ideas underpinning public service media developed over the last century in the context of governments seeking to strengthen their societies' democratic and representative channels of communication as well as means of reliable public information provision and cultural exchange \cite{scannell1991social,born2005uncertain,seaton2021bbc,hendy:psb}. Although these normative principles originated in national democratic polities, they have also been applied transnationally, for example in the European Union through the auspices of the European Broadcasting Union. They are therefore not limited in their purview to national media systems~\cite{jakubowicz2007public,jakubowicz2010psb,dobek2010comparative,donders2011public}. A common misconception is to equate PSM systems with state-controlled media; however, as media institutions and ecologies created for the purposes mentioned, and bolstered by regulatory frameworks, they are designed to be independent of the state, to exhibit a certain autonomy and political impartiality, and to be publicly accountable \cite{born2005uncertain}---although in reality this status may be fragile or imperfectly achieved.

A substantial body of research in media and political theory has identified the normative principles informing PSM systems and how these systems function as a communicative infrastructure for democratic societies. Central among those principles are universality (or commonality), diversity, and citizenship~\cite{scannell1989public,raboy1996public,born:21stcentury,born:ecology,born-prosser:psm-principles,born:2006digitilizing}. We consider this triad particularly relevant for recommender systems delivering cultural content, since together they answer calls in democratic media and political theory for digital media systems to enhance cultural citizenship~\cite{born:2006digitilizing}. The concept of cultural citizenship has become foundational for democratic political theories in the last two decades; indeed, ``one of the striking developments in recent political discourse has been the increasing confluence of culture and citizenship'' \cite{delanty:cultural-citizenship}. Cultural citizenship has been defined as a fourth stage of citizenship that responds to recognition of the social transformations and challenges posed by globalization, increased migration, the growing heterogeneity of the populations of nation states, and the intensification of identity politics among subaltern and marginalized groups~\cite{rosaldo1994cultural,miller2001introducing}.\footnote{In Marshall's classic sociological account of the historical emergence of citizenship \cite{marshall:citizenship-and-social-class}, he divides it into three stages or `elements'---civil, political, and social \cite{evers:social-policy-and-citizenship}. Theorists of cultural citizenship conceive of it as a fourth stage.} Given these profound changes, cultural citizenship draws attention to a `new domain of cultural rights [involving] the right to symbolic presence, dignifying representation', and `the maintenance and propagation of distinct cultural identities'~\cite{pakulski1997cultural}. Hence, for theorists of cultural citizenship, 'cultural pluralism is viewed as something which enriches rather than threatens the fabric of society' \cite[p. 61]{delanty:cultural-citizenship}.  In this light it becomes clear that, in order to promote cultural citizenship, PSM organizations and other democratic channels of cultural production and distribution have a responsibility to curate and disseminate a plurality of cultural content with the intention of stimulating both intercultural and intracultural dialogue, as well as the acceptance of, and respect for, cultural diversity \cite{born:2006digitilizing,born2012mediating}. In this way PSM and other democratic media can act both as a force `for social cohesion and integration' and as a forum for pluralistic cultural experience and exchange among those many groups and communities that coexist and interact in democratic societies~\cite{jakubowicz2007public}.

Music, movies, and literature, as expressive media, add further dimensions to these ideas. Some political theorists argue that the dialogical mechanisms of democratic pluralism should not be confined to the classic concerns of public sphere theory---information, reason, and cognition---but should also engage matters of identity and affective experience. Hence, the political philosopher Martha Nussbaum draws attention to emotion as a basic component of ethical reasoning, arguing that a compassionate citizenry depends on access to pluralistic cultural repertoires that engage audiences' emotions and thereby enhance their capacity for mutual recognition, empathy, and toleration. For Nussbaum such processes are essential for the well-being and development of democratic societies~\cite{nussbaum2003upheavals}. Arguably, then, cultural citizenship is the principal form for the exercise of citizenship in the multicultural societies characterizing the contemporary world. If we take seriously the role of mediated cultural content, such as that curated by recommender systems, in influencing users' tastes and thereby conditioning the wider public culture, then, by analogy with the concern in democratic theory with the formation of an educated and informed citizenry, we might add a concern with the formation of a culturally mature and aware, culturally pluralistic citizenry~\cite{born-prosser:psm-principles,parekh2000parekh}. In this sense, digital platforms distributing cultural content---such as music, movies, and literature---can be understood as primary `theaters' for contemporary pluralism and consequently bear an obligation to provide a diversity of cultural experience. As Stuart Hall, a leading critical race theorist, has noted, `The quality of life for black or ethnic minorities depends on the whole society knowing more about the “black experience”'~\cite{hall1993}, an experience that can be grasped most compellingly through access to the diverse riches of black cultural production---whether music, movies, or literature. Platforms curating cultural content therefore have the capacity, and arguably the responsibility, to play a vital role in fostering cultural citizenship---itself a precondition for the processes of ethical, social, and cultural development that underlie the general condition of citizenship~\cite{born:2006digitilizing}.

Both universality or commonality---that is, the provision of common cultural experiences---and diversity of cultural experience, or exposure to diverse cultural content, are therefore essential to the strengthening of cultural citizenship. As Georgina Born argues, both `mutual cultural recognition and the expansion of cultural referents… are dynamics essential to the well-being of pluralist societies. But this does not obviate the need also for integration---for the provision of common [cultural] experience and the fostering of common identities'~\cite{born:2006digitilizing}. Scholarship on these matters emphasizes, further, that implementing principles like universality (commonality), diversity, and citizenship require `alternative success metrics… focused on [media systems'] impact on democracy' and which address users `as citizens and not just… as consumers'~\cite{unterberger2021public}. Such metrics will enable democratically-oriented media and platforms to adapt to the present by advancing `cultural citizenship and the needs of the digital society'~\cite{jakubowicz2007public}. Recommender system design intended to strengthen cultural citizenship therefore requires us to implement universality---via a commonality metric---and to deliver diversity of cultural experience, a challenge to which we turn now.

\subsection{Human editing, value communities, and recommender systems as sociotechnical assemblages}
\label{sec:motivation:editors}

Given the importance of pluralistic cultural experience in strengthening cultural citizenship, a core challenge for this research is the need to boost the diversity of cultural content to which a population of users is exposed by recommendation. Unlike existing ideas of diversity employed in the recommender system literature, we consider that diversity for cultural items such as movies, music, or literature can be conceptualized in a range of ways. They include, first, diversity of content in terms of artistic and cultural expression, which can be equated with the need to ensure that a range of genres are present as well as intra-generic differences, generic margins and niches. And second, diversity of the source or producer of the content, according to region, territory, or culture of origin as well as under-represented demographics among producers of the content (musicians, filmmakers, writers). The two---diversity of content and of source---are potentially related, in that greater diversity of source or producer is likely to favor, although it does not guarantee, greater diversity of content. However, judgments about what kinds of content and source diversity are desirable are intrinsically context-dependent and culturally-dependent. In this sense they necessitate human editorial processes that draw on knowledgeable and communally-validated categorizations---whether of the subtleties of demographic categories, or of the complex contours of cultural genres. 

A key assumption in our work is therefore that human editors must be involved in these judgments, and that their role is to reflect on the diversity of a recommender with respect to a given category or categories by drawing on insights generated by a larger `value community' knowledgeable about cultural expressions and their social conditions~\cite{born2010social}. 
By value community we refer to the existence of communities sharing cultural interests and tastes, among them genre communities, who broadly embody an evolving consensus about the cultural interests or genres they enjoy and their relationship to categories of social identity, and about which members have varying degrees of expertise. The consensual judgments of value emerging from a value community are, then, relational, and, as Bourdieu suggests, they will inevitably encompass a lively and shifting dissensus within the consensus \cite{bourdieu:systems-of-education}. The human editors we envisage therefore act as conduits for these larger communities of interest and judgement, and their judgments are legitimised and validated by this relationship.

The aim is to achieve a diverse mix of content and sources that appeals beyond personalization, and that avoids the risks of employing reified models of both identities and genres. Editors' judgments, moreover, will necessarily evolve over time and will be repeatedly replenished by evaluating (via a commonality metric) the performance of the recommendation of the diverse categories selected across a user population. It is the resulting universal promotion of a plurality of cultural experiences, relative to a given social context and cultural situation, that is likely to cumulatively enhance cultural citizenship; it may also foster progressive cultural and social change.  

A related conceptual step is necessitated by the key role we are proposing for editors responsive to wider value communities. It is to expand how we think about `recommender systems' to include human editors, the value communities validating their knowledge, and user populations (or audiences). In this light we propose that recommender systems can be productively conceived as sociotechnical assemblages that include the social knowledge and social labour that go into the processes described. As Seaver puts it, `algorithms are not autonomous technical objects, but complex sociotechnical systems', and `while discourses about algorithms sometimes describe them as “unsupervised,” working without a human in the loop, in practice there are no unsupervised algorithms. If you cannot see a human in the loop, you just need to look for a bigger loop'~\cite{seaver2018should}. Designating recommender (or algorithmic) systems as sociotechnical assemblages implies, then, that these `technologies are embedded in the social context that produces them'~\cite{sapignoli2021anthropology}.

\section{Conceptualizing diversity in relation to cultural citizenship, and implementing it in recommender systems}
\label{sec:relatedwork}

If diversity of cultural experience is a precondition for enhancing cultural citizenship, then the question is how diversity should be conceptualized in relation to recommender systems in order to achieve this end. As discussed above, diversity of content and diversity of source are perhaps the most obvious vectors of diversity. But it is certainly possible to imagine additional forms of diversity in relation to recommender systems focused, for example, on diversity of consumption experience and of user controls. This might include the potential to design diversity into the navigational architecture of recommendation by avoiding `similarity' and promoting difference; or by offering controls to users that endow the algorithm with greater legibility and increase users' agency to pursue diverse pathways through a given recommendation space. 

Yet although these and other approaches to diversity might be significant for design, the RecSys community has mainly addressed diversity in terms of promoting either diversity in some abstract space (e.g. a vector space) or through fairness measures, approaches that equate in some ways to what what we have called diversity of content and/or of source. Existing work, conceptualizing diversity in this way, even if it is concerned with and sensitive to the broader social role of cultural content, does not adequately support the rich set of goals system designers might have and the values they might want to implement through design. Typically, diversity metrics are limited to the goal of capturing the variety of content \textit{within} a recommendation list; they may consider categorizations of the content, distances in a latent space, or simply how many different items are recommended \cite{schedl2015, Yucheng2018, anderson2020, Han2017ASO}. Aligned with the goals of personalization, the formulation of these diversity metrics sometimes optionally consider the relevance of the content for users, assuming that what a user consumed in the past indicates what they will still be interested in and that recommendations should be limited to such categories. And while related novelty metrics measure the newness of items or categories of recommendations, they are still individualized, and they are also agnostic about \textit{what type} of content is new to the user. 

In a similar way, to evaluate fairness means to be concerned with increasing diversity by seeking to redress the problematic under-representation of certain categories of source and content. But these approaches also tend to adhere to fixed and pre-given definitions of genres and identities, in this way risking the reification of those categories and untethering them from processes of community validation of the kind we advocate in Section \ref{sec:motivation:editors}. Existing work on fairness addresses specific topics around increasing biases as well as the under-representation of particular groups \cite{Ekstrand2022}.  Provider fairness metrics typically consider how many different groups of content providers appear in recommendations and assume a given distribution that it is desired to match. Consumer fairness metrics consider disparate treatments of the system in relation to different groups of consumers. Recent research has proposed more general multi-stakeholder fairness metrics, acknowledging the impact recommender systems have on different groups of individuals \cite{milano2021ethical,burke2016towards,sonboli2022}.

A standard argument is that by reflecting biases embedded in the datasets, recommender systems create a feedback loop reinforcing such biases \cite{chaney:recsys-feedback-loops,mansoury2020feedback}. The loop can be identified in popularity bias, which may reflect a mainstream bias in cultural domains. \citet{slaney2006measuring} studied this propensity in music playlists and theorized diversity as a key criterion for user satisfaction, providing music discovery for users who ``do not want to listen to the highest rated song'' within a system ``over and over again.'' \citet{celma2008hits} noted similarly the tendency to reinforce ``popular artists, at the expense of discarding less-known music.'' Both in provider and consumer fairness, recommender systems have been shown to reproduce or exacerbate wider conditions of cultural and social discrimination against certain social groups.

Existing work on fairness addresses specific topics around increasing biases as well as the underrepresentation of particular groups \cite{Ekstrand2022}. Provider fairness metrics typically consider how many different groups of content providers appear in recommendations and assume a given distribution that it is desired to match. On the other hand, consumer fairness metrics considers disparate treatments of the system in relation to different groups of consumers. Recent research has proposed more general multi-stakeholder fairness metrics, acknowledging the impact recommender systems have on different groups of individuals \cite{milano2021ethical,burke2016towards,sonboli2022}. 

On the consumer fairness side, \citet{mansoury2020feedback} show that biases against minority groups of users can be reinforced in movie recommender systems as they reinforce user choices “through different iterations of users interaction” with the system. In their study, they highlight stronger bias amplification in recommendations for female users. \citet{shakespeare2020exploring} evaluate in the music domain how gender bias, rooted ``in cultural practices historically related with socio-political power differentials,'' can be ``propagated by CF-based recommendations'' based on user ratings. In similar research on movie recommendation, \citet{lin2019crank} demonstrates the propagation of users' gender biases, arguing that ``recommendation algorithms generally distort preference biases present in the input data and do so in sometimes unpredictable ways.'' \citet{ekstrand2018all} demonstrate how popularity and demographic biases in both music and movie recommendation tend to affect user utility grouped by age and gender; they show that the models tend to perform better for male users and vary significantly across age groups. \citet{kowald2020unfairness} studies how popularity bias may lead to unfair treatment of users with little interest in popular items in the context of music recommendations. \citet{kowald2021support} show that recommendations' accuracy varies for different groups of users depending on their openness towards beyond-mainstream music listened. \citet{lesota2021analyzing}  studies how multiple recommender systems with different levels of popularity bias may affect users of different genders differently.
 
On the provider fairness side, \citet{epps2020artist} shows for music recommendation that female and non-binary artists can be under-represented, affecting users' listening behaviors. They argue that ``higher proportions of female artists in recommended streaming is predictive of higher proportions of female artists in organic streaming.'' \citet{ferraro2021break} also reveal an imbalance of exposure in music recommendations between female and male artists, and a tendency to confirm the feedback loop that moves male artists to the top. \citet{ekstrand2018} observe, with respect to books, that ``there are efforts in many segments of the publishing industry to improve representation of women, ethnic minorities, and other historically underrepresented groups.'' Yet they argue that the recommender systems they analyzed tend to propagate disparities present in user profiles. This tendency in the book domain was also confirmed by \citet{saxena2022exploring}.

Overall, it is striking that the various forms of bias shown by the under-representation of cultural content with respect to gender, race, class, and region (i.e., diversity of source or provider) correspond to wider core-periphery dynamics and geographical inequalities in the cultural industries \cite{tofalvy2021splendid,camposinequalities, verhoeven2019re}. It seems that recommender systems often mirror these inequalities, promoting Western-centric popular cultural content, in the English language, released by major producers \cite{voit2021bias}. Increasing diversity of source and producer, both as an issue of equity in itself and as it bears on diversity of content, is therefore a huge hurdle in achieving recommender systems oriented to enhancing cultural citizenship.

\section{Measuring Commonality}

Criticisms of recommender systems and machine learning systems for reproducing such forms of bias and discrimination have been met by personalized recommender systems interventions aimed at redressing bias at the level of individual users. In light of our discussion of the importance of addressing the cumulative cultural and social effects of recommender systems, we contend that it is also crucial to identify means of counteracting bias and enhancing the diversity of source and content offered across populations of users. We propose measuring common experiences of diversity at the aggregate level. Assuming a democratic media environment, we seek to evaluate whether a recommender system  contributes to the strengthening of cultural citizenship by systematically promoting diversity of source and content within a given type of cultural content (in our experiments, movies, music, and literature). In this way, evaluation has the potential to assist in counteracting sexist and racist biases and the neglect of non-Western and non-mainstream content across a user population. This also provides a means of evaluating the extent to which a given recommender system is contributing to the kinds of wider cultural changes called for by anti-racist and feminist critics as well as by those sympathetic to criticisms of existing recommender systems.

\subsection{Metric Definition}\label{sec:commonality}
Recall that we are interested in measuring the extent to which users, in response to algorithmic recommendations, gain a shared familiarity with a diverse set of content.  This requires us to define
\begin{inlinelist}
    \item which items or categories of items should be emphasized,
    \item how we quantify familiarity,
    \item how we quantify a shared familiarity, and
    \item how we aggregate multiple categories of commonality.
\end{inlinelist}

\subsubsection{Selecting Categories}
\label{sec:metrics:editors}
The promoted categories, we suggest, will be identified and curated by editors in a relevant field seeking to promote a plurality of cultural content in the service of strengthening cultural citizenship.  We contrast this with statistical methods for selecting under-represented categories (e.g. \cite{mehrotra:fair-marketplace}), which may surface under-represented content misaligned  with the goals of enhancing diversity of cultural experience across a user population, and at the same time enhancing their common experience of that diversity. As described in section \ref{sec:motivation:editors}, editors make curatorial decisions drawing on insights generated by their knowledge about cultural expressions and social conditions with the goal of achieving a diverse mix of content and sources that appeals beyond personalization, and that avoids the risks of employing reified models of both identities and genres. These editors may opt to promote, for example, movies by female directors or those produced for non-Western markets. 

Given the large body of criticism of bias and lack of fairness in the RecSys literature, and for the purpose of testing the commonality metric, in what follows we chose to work experimentally with widely recognized under-represented categories of source or producer in the three chosen media (movies, music, and literature). The under-represented categories come in three broad clusters, which are: female and non-binary gendered providers, artists or authors; independent production; and non-Western sources. At the same time, boosting diversity by promoting these under-represented source and producer categories bears directly on, and is very likely to increase, the diversity of content in each case. However, it is important to point out that the approach and the principles set out in this article can be applied in alternative ways, employing different categories and boosting different vectors of diversity. The distinctive facet of our work is not so much the attempt to redress specific kinds of under-representation -- although we are certainly concerned with this challenge both in itself and as a key component of the goal of enhancing cultural citizenship. Rather, it is the attempt to bind such an intervention to larger normative ambitions  (strengthening cultural citizenship), and to find means of evaluating the effects of this intervention (that is, increasing the diversity of cultural experience) not just on individuals but universally, across populations of users.

\subsubsection{Measuring Familiarity}
\label{sec:metrics:familiarity}
In order to measure familiarity, we make the simplifying assumptions that all users 
\begin{inlinelist}
    \item begin their recommendation session with the same background in the relevant categories (Section \ref{sec:metrics:editors}), and
    \item engage with exactly one ranked list of recommendations.
\end{inlinelist}
We make these assumptions to clarify our exposition and  extensions to heterogeneous backgrounds and multi-turn recommendations are left for future work.  

A user's familiarity with a category after having interacted with a ranked list of recommendations can be computed using existing evaluation methods.  If a user has interacted with the $k$ items, then the familiarity can be estimated by computing the fraction of all items in the category present amongst the $k$ items.  Let $\ranking_{\user}$ be a ranking of $\numitems$ items from the catalog $\catalog$ for user $\user\in\users$.  If a user scans linearly from the first ranked item downward, we can measure familiarity as the \textit{recall} of items in category $\category$ at position $k$,
\begin{align}
    \label{eq:recall}
    \recall(\ranking_\user,k,\category)&=\frac{|\ranking_{\user,:k}\cap \catalog_{\category}|}{|\catalog_{\category}|}
\end{align}
where $\catalog_{\category}$ is the set of items labeled with category $\category$.  

Although we could use a fixed cutoff $k$, this may not capture users that terminate their scan of the list before or after the $k$th item.  A user browsing $\prob(k)$ model provides us with a distribution over possible stopping positions.  Specifically, in our experiments, we adopt the browsing model used in rank-biased precision \cite{moffat:rbp},
\begin{align}
    \label{eq:rbp}
    \prob(k)&=(1-\gamma)\gamma^{k-1}
\end{align}
where the patience parameter $\gamma\in(0,1)$ controls how deep into the ranked list the user is likely to progress, regardless of relevance.

Combining Equations \ref{eq:recall} and \ref{eq:rbp}, we can compute, given a ranking $\ranking_\user$, the familiarity of $\user$ with category $\category$ as the expected recall,
\begin{align}
    \label{eq:erecall}
    \prob(\familiarity_{\user,\category}|\ranking_\user)&=\sum_{i=1}^{\numitems}\prob(i)\recall(\ranking_\user,i,\category)
\end{align}
where $\familiarity_{\user,\category}$ is a binary random variable indicating that the user is familiar with the category.  
\subsubsection{Commonality}

The  \textit{commonality} of a system captures the probability that \textit{every} user simultaneously gains familiarity with the editorially-selected categories. 

Given a set of editorially-selected categories $\categories$, we can compute the commonality of a system with respect to a single category $\category\in\categories$ as the probability that every user has become familiar with $\category$ under the system's ranking,
\begin{align}
    \commonality_\category(\rankings)&=\prob(\familiarity_{1,\category},\ldots,\familiarity_{\numusers,\category}|\rankings)\nonumber\\
    &=\prod_{\user\in\users}\prob(\familiarity_{\user,\category}|\ranking_\user)    \label{eq:commonality}
\end{align}
Using the joint probability of familiarity over all users emphasizes the importance of the public in our conception results in a metrics that degrades quickly, even if few (or one) user whose category recall is low.   In practice, to address numerical precision issues,  we use the logarithm of commonality, which is rank equivalent with commonality.  

\subsubsection{Aggregation}
While some system designers may be comfortable analyzing commonality disaggregated by category, many will want a summary of commonality across groups.  This may be due to convenience (e.g. a leaderboard) or out of a desire to measure robust performance across categories, in the interest of promoting a shared plurality of cultural content.

As such, we developed an aggregated commonality metric, summarizing performance across categories. Although we could aggregate per-category commonality using an arithmetic mean, this might be instable due to different category sizes and sensitive metric (Equation \ref{eq:commonality}).  Instead, we adopt Borda's rank aggregation method.  Assume that we have a set $\systems$ of systems, each associated with a set of per-user rankings $\{\rankings^s\}_{s\in\systems}$.
We begin by, for each category $\category\in\categories$, generating a system ranking according to $\commonality_\category(\rankings^s)$.  We then assign the top-ranked system with a value of $1$, the second-ranked system with a value of $2$, down to $|\systems|$ for the last-ranked system.  Aggregating these `votes' across categories results in a final system score, where lower values are better.

\subsection{Mathematical Analysis}
\label{sec:metrics:analysis}
In Section \ref{sec:relatedwork}, we discussed prior approaches to measuring the diversity and fairness of recommender systems in the context of cultural citizenship.  Given that our commonality metric is developed with cultural citizenship in mind, we turn to contrasting the specific mathematical differences between commonality and prior metrics.

\subsubsection{Aggregation}
The fundamental conceptual shift from individualized to collective metrics is reflected in our adoption of the joint probability of user events.  As a simple contrast, consider normalized discounted cumulative gain ($\ndcg$) \cite{NDCG}.  When computing the aggregate metric, we look at the sample mean over users,
\begin{align*}
    \E_{\users}[\ndcg(\rankings)]&=\frac{1}{|\users|}\sum_{\user\in\users}\ndcg(\ranking_\user)
\end{align*}
where $\rel_\user$ be the set of items relevant to user $\user$.  In this case, we can see that $\ndcg$ sums the metric value across users while commonality (Equation \ref{eq:commonality}) multiplies familiarity across users.  This results in a metric that is much more sensitive to supporting collective familiarity and shared cultural experiences.  

As a general case, diversity and fairness metrics operate similarly.  We adopt an individualized metric, compute its value for a user (e.g. answering `how fair/diverse is this ranking for this user?'), and then compute the sample mean.  As a result, we can imagine situations where the lack of diversity or fairness for some users is `compensated for' by users whose recommendations are more diverse or fair.  

One exception to this is fairness metrics based on measuring the Kullback-Leibler divergence between the distribution of categories in recommendations from a uniform distribution over categories.  Let $\theta$ be the distribution of exposure over recommended groups and $\theta^*=\frac{1}{|\categories|}$ a uniform distribution over groups \cite{kirnap2021estimation}.  We can reduce the sample mean of this metric to a rank equivalent quantity based on the sum of group joint probabilities,
\begin{align*}
    \E_{\users}[\dexpkl(\rankings)]&=\frac{1}{|\users|}\sum_{\user\in\users}\kld(\theta^*\|\theta_\user)\\
    &=\frac{1}{|\users|}\sum_{\user\in\users}\sum_{\category\in\categories}\frac{1}{|\categories|}\log\left(\frac{\frac{1}{|\categories|}}{\theta_{\user,\category}}\right)\\
    &\rankeq-\sum_{\user\in\users}\sum_{\category\in\categories}\log\left(\theta_{\user,\category}\right)\\
    &=-\sum_{\category\in\categories}\log\left(\prod_{\user\in\users}\theta_{\user,\category}\right)
\end{align*}
In this case, we can see that, like commonality,  $\dexpkl$ includes a product of per-user metrics. However, there are two slight differences.  First, the metric being multiplied is the relative exposure of a category in the user's ranking as opposed to the familiarity.  While these may sometimes be correlated, there are certainly situations where we might observe high $\theta_{\user,\category}$ and a low $\familiarity_{\user,\category}$, meaning that $\dexpkl$ would be inappropriate for measuring the shared familiarity.  Second, the aggregation of joint metrics in $\dexpkl$ uses a simple sum aggregation, which is possible, in part, because $\theta_{\user,\category}$ is calibrated across groups while $\familiarity_{\user,\category}$ may not be (i.e. differences in sizes of categories may lead to different ranges of empirical values).  Note that this observation may be unique to using the Kullback-Leibler divergence, which includes a logarithmic term.  

Another way to interpret category commonality is as the geometric mean of the recall of a category,  connecting it to geometric mean average precision \cite{roberston:gmap}. In the context of utility metrics, \citet{valcarce:recsys-ranking-metrics-journal} experiment with geometric mean performance, finding that it is more robust than the arithmetic mean when dealing with samples of users.  We will return to this observation in Section \ref{sec:results:missing-users}.

\subsubsection{Category Metric}
A second difference between commonality and prior fairness, diversity, and novelty metrics is in the category-level metric.  

Fairness metrics tend to emphasize divergence from some reference distribution of categories \cite{kirnap2021estimation}.  In the previous section, we saw the example of $\dexpkl$, where, when using the uniform distribution over categories as a reference, the aggregated metric reduces to the magnitude of categories in the recommendations.  These per-user quantities capture the the \textit{presence} of the category as opposed to the \textit{comprehensiveness} of the category (i.e., recall).  Even in the case of non-uniform reference distributions, exposure distributions are normalized in such a way that any recall information is removed, implying that, while similar in form, fairness metrics are mathematically measuring a different phenomenon.  

While fairness metrics capture the divergence of category exposure from some reference distribution, diversity metrics measure the support of the exposure distribution in some space.   Consider the expected intra-list distance ($\eild$).  Assume that we consider all recommended items,
\begin{align*}
    \E_{\users}[\eild(\rankings)] &= \frac{1}{|\users|}\sum_{\user\in\users}\sum_{i=1}^{\numitems}\sum_{j=i+1}^{\numitems}\prob(i)\prob(j-i)\delta(\ranking_{\user,i},\ranking_{\user,j})
\end{align*}
where $\delta$ is a linear distance function between items.  If each item belongs to only one category, then  we can set $\delta(i,j)=1-\sum_{\category\in\categories}\ind(i\in\catalog_{\category})\ind(j\in\catalog_{\category})$ and derive a rank-equivalent metric,
\begin{align*}    
    \E_{\users}[\eild(\rankings)] &\rankeq-\sum_{\category\in\categories}\sum_{\user\in\users}\underbrace{\sum_{i=1}^{\numitems}\sum_{j=i+1}^{\numitems}\prob(i)\prob(j-i)\ind(\ranking_{\user,i}\in\catalog_{\category})\ind(\ranking_{\user,j}\in\catalog_{\category})}_{\text{exposure of items in $\catalog_\category$}}
\end{align*}
From this, we can see that, like commonality, diversity computes the exposure of items in a category. Like fairness metrics, diversity metrics measure presence as opposed to comprehensiveness.  

In terms of novelty, take the expected profile distance ($\epd$) metric \cite{vargas2011}.  Assume that we consider all recommended items and relevant items in the users profile,
\begin{align*}
    \E_{\users}[\epd(\rankings)] &= \frac{C}{|\users|}\sum_{\user\in\users}\sum_{i=1}^{\numitems}\prob(i)\sum_{r\in\mathcal{R}_{\user}}\delta(\ranking_{\user,i},r)
\end{align*}
where $C$ is a constant and $\delta$ is a linear distance function between items.  If each item belongs to only one category, then  we can set $\delta(i,j)=1-\sum_{\category\in\categories}\ind(i\in\catalog_{\category})\ind(j\in\catalog_{\category})$ and derive a rank-equivalent metric,
\begin{align*}    
    \E_{\users}[\epd(\rankings)] &\rankeq-\sum_{\category\in\categories} \sum_{\user\in\users}\underbrace{\sum_{r\in\mathcal{R}_{\user}}\ind(r\in\catalog_{\category})\sum_{i=1}^{\numitems}\prob(i)\ind(\ranking_{\user,i}\in\catalog_{\category})}_{\text{\textit{new} exposure of items in $\catalog_\category$}}
\end{align*}
From this perspective--and with this distance function--we can see that, for each category, the metric measures magnitude of exposure of items in that category \textit{for users who have already engaged with an item in that category}.  For example, in the movie domain, taking the category of West African film, if a user has already positively rated at least one movie in that category, this metric would measure how many more West African films are recommended. 

\subsection{Empirical Analysis}
\label{sec:experiments}

Although we synthesized concepts from the PSM literature and the evaluation literature to develop our commonality metric, we are interested in understanding its empirical behavior as an evaluation metric.  To this end, we explored the following,
\begin{inlinelist}
    \item correlation with existing metrics,
    \item robustness to missing labels,
    \item generalizability from population samples, and
    \item optimizability.
\end{inlinelist}
These analyses use a fixed experimental setup consisting of multiple, publicly available datasets, which we use to compare commonality with existing metrics.  Our analyses focus on the behavior of the commonality metric under different possible editorial policies.  So, while a production system would employ editors who act as conduits for value communities, we select categories such that they are representative in size and representation to what we might expect from a human editor.

\subsubsection{Data}
We consider three recommender system domains dealing with cultural content: movies, music, and literature.  For each dataset, in addition to publicly available data, we selected categories (i.e. $\categories$) based on their historic under-representation in order to assess the behavior of our metric.

\paragraph{Movies}
We use the movielens-1m dataset, which contains 1,000,209 ratings of approximately 3,900 movies from 6,040 users from the movielens platform. Using a separate dataset\footnote{https://www.themoviedb.org/}, we augmented the movielens movies with metadata including country of production, gender of the director, original language, and keywords collected from the movie's description.  For this dataset, we used rankings from multiple recommendation systems prepared by \citet{valcarce2018}. Following the method described by the authors we converted to binary relevance labels considering ratings of 4 and 5 as relevant. 
We selected categories of movies that are typically under-represented by movie recommender systems. Specifically, we consider female directors (under-representation by gender); independent film (under-representation by industry sector); and several sources of non-Western film (under-representation by geographical and linguistic inequality).  We use categorical gender data, acknowledging the limitations of this framing \cite{hoffmann:fairness-failure}.  For geographic categories, we use the country of production for the following regions of the world: 
South America, Central America, North Africa, South Africa, West Africa, Mid Africa, Southeast Asia, South Asia, Western Asia, Central Asia and East Asia.
We consider, broadly, non-English language movies as a separate category.  
And, finally, we use keywords to create categories with selected movies whose categories contain ``independent films'', ``LGBT'', and ``transgender''. We manually checked whether these keywords can be trusted to represent specific identities. 

\paragraph{Music}
For music, we use Last.fm-2b dataset, which is the largest dataset containing users' listening events for 120,000 Last.fm users, and over two billion listening events. We enriched this dataset with additional information
of the artists collected from MusicBrainz.org, a large collaborative database of music information. 
From this dataset we only considered interaction between the years 2013 and 2020. After removing items that were listened to by fewer than fifteen users and users that listened to less than fifteen items, the resulting dataset had 18,711 users and 28,341 items.  From these items a total of 2,712 belong to at least one category. 
We selected eight categories that are related with regions of the world of non-western countries from the artists' locations (North Africa, East Africa, Middle Africa, South Africa, West Africa, Middle East, Central Asia and Southeast Asia). We additionally selected two categories based on artists' gender information collected from Musicbrainz, these two categories are female artists and non-binary artists. We acknowledge that the category of  female artists represent a group  much larger than the other categories but, as we have reviewed, their are still under-representend or suffer through both industrial and recommender biases when compared to male artists. 
We trained 12 recommender systems for the last.fm data using the Elliot library. We use models based on MF2020, NeuMF, RP3beta, BPRMF and iALS trained of both binarized and original input; plus two baselines: Popularity and Random. 

\paragraph{Literature}
For literature, we use the Librarything dataset, containing user book ratings. We use a subset of the dataset containing 7279 users and 37232 items. We collected information for these books from Librarything.com. From the information we selected the following categories: Africa, Asia, Latin-america, Middle-east, Environment, Female, LGBT, Independent and Multicultural \& Race. A total of 8171 books correspond to at least on category.
For this dataset, we also used rankings from multiple recommendation systems prepared by \citet{valcarce2018}. Following the method described by the authors we first converted the original ratings to a scale of 1-5 and then converted to binary relevance labels considering ratings of 4 and 5 as relevant.

\subsubsection{Baseline Metrics}
\label{sec:baseline-metrics}
For all the experiments, we compare commonality against three classes of metrics: utility metrics, diversity metrics~\cite{vargas2011}, and fairness metrics \cite{canamares:offline-recsys-eval}.  

We measure utility-focused properties using precision ($\precision$), recall, ($\recall$), and $\ndcg$.  
We measure fairness across categories $\categories$ using  disparate exposure ($\dispexp$) \cite{gomez2021} and the divergence family of metrics ($\dexpabs$, $\dexpsq$, $\dexpkl$) using the probability of exposure to categories \cite{kirnap2021estimation}.  
We measure diversity of categories $\categories$ using $\andcg$ and $\iaerr$ and novelty using Expected Intra-List Distance ($\eild$) and Expected Profile Distance ($\epd$) with distances based on genre representations of items.

\subsubsection{Correlation with Existing Metrics}
\label{sec:metrics:results:correlation}
In our first analysis, we were interested in understanding the correlation between commonality and existing metrics (Section \ref{sec:baseline-metrics}).  Observing consistently high correlations between commonality and existing metrics across domains would suggest redundancy, reducing the need for a new metric.  We measure this correlation across three different editorial regimes for selecting items within each category (i.e. which subset of $\catalog_\category$).  In the first condition, we assume that editors select \textit{all} items in $\catalog_\category$.  For example, if items authored by women were a broad category of interest, an editor in this condition would be interested in promoting a comprehensive familiarity with all items authored by women.  In this condition,  larger categories, while sometimes more likely to be recommended naturally (if the category is already popular), will be more difficult to achieve high familiarity with, even when explicitly programmed; smaller categories, on the other hand, will, by chance, have a lower probability of exposure but, with explicit programming, can reach high familiarity.  In part to address this, we consider a second regime wherein an editor may downsample items from the largest category.  Our final regime considers downsampling items from all categories until they have equal size.  In this section, our analyses help us understand inter-metric correlation as a function of different editorial conditions.  

In order to compare metrics, we compute the Kendall's $\tau$ correlation between system rankings according to commonality and existing metrics.

\begin{table}[t]
  \caption{Full Category Selection. Correlation between commonality utility, fairness, and diversity metrics.  Kendall's $\tau$ between rankings of runs with Bonferroni correction to correct for multiple comparisons (bold: $p<0.05$). }\label{tag:borda-comm-correlation}

  \centering
          \centering
          \begin{tabular}{lccc}
            \hline
            &  movies	&	music	&	literature\\
            \hline
            \textbf{Utility}\\
            $\precision$	&	-0.119	&	\textbf{0.512}	&	0.053	\\
            $\recall$	&	-0.138	&	\textbf{0.574}	&	0.043	\\
            $\ndcg$	&	-0.205	&	\textbf{0.450}	&	0.072	\\
            \textbf{Fairness}\\
            $\dispexp$	&	\textbf{0.721}	&	0.326	&	\textbf{0.763}	\\
            $\dexpkl$	&	\textbf{0.616}	&	\textbf{0.698}	&	0.062	\\
            $\dexpsq$	&	\textbf{0.348}	&	\textbf{0.636}	&	\textbf{-0.647}	\\
            $\dexpabs$	&	\textbf{0.339}	&	\textbf{0.605}	&	\textbf{-0.647}	\\
            \textbf{Diversity and Novelty}\\
            $\andcg$	&	-0.062	&	\textbf{0.512}	&	0.254	\\
            $\iaerr$	&	-0.100	&	\textbf{0.512}	&	0.254	\\
            $\eild$	&	\textbf{0.730}	&	\textbf{0.853}	&	\textbf{0.782}	\\
            $\epd$	&	\textbf{0.702}	&	\textbf{0.822}	&	\textbf{0.763}	\\
            \hline
  \end{tabular}
\end{table}

\paragraph{Full Category Selection}
In our first analysis, we compare the correlation when using an editorial policy that selects all items in a category for promotion.  Our results  (Table \ref{tag:borda-comm-correlation}) indicate that none of our utility and fairness metrics show a strong consistent correlation with commonality.  While some domains show stronger correlations for some of these metrics, there is no evidence that commonality is redundant with these metrics.  In terms of diversity and novelty, both $\eild$ and $\epd$ show stronger correlation with commonality consistently across domains.

\paragraph{Downsampling the dominant category}
In order to understand the relation between commonality and previous metrics under different editorial policies, we now look at how selecting a subset of items in the larger category would affect the rank correlations. 
In this analysis, we progressively remove category labels for items from the dominant category, simulating a policy that de-emphasizes familiarity with the comprehensive set of items in a category. We randomly downsample the dominant category to percentages of the original size, including (10, 30, 50, 70 and 90\

\begin{figure}[t]
\centering

\includegraphics[width=\textwidth]{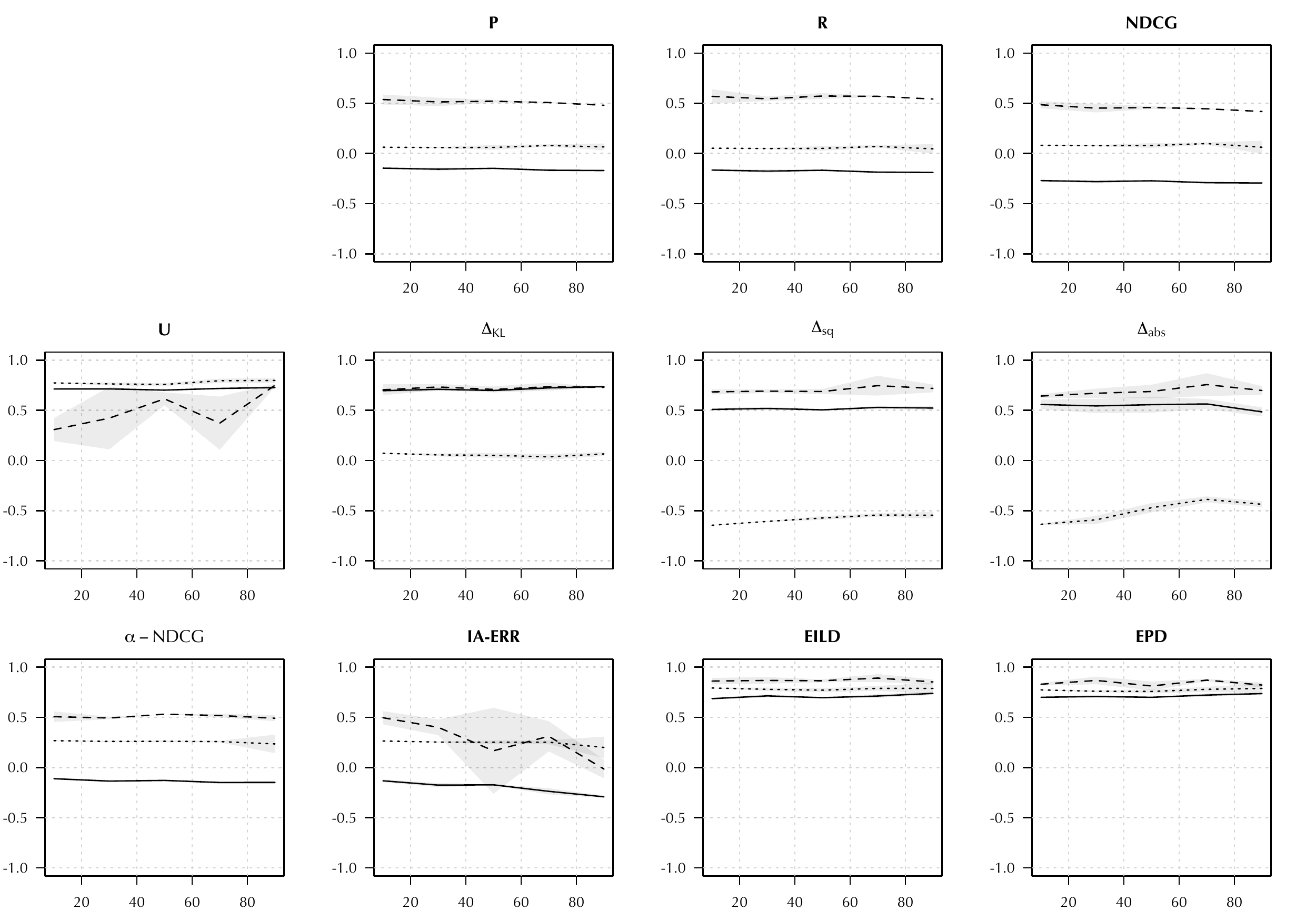}
\caption{Correlation with commonality when downsampling the dominant category.  Horizontal axis (all plots): percentage downsampled. Vertical axis (all plots): $\tau$ correlation with commonality. Top row: Utility metrics; middle row: fairness metrics; bottom row: diversity metrics. Solid line: movies; dashed line: music; dotted line: literature. Lines show mean across five trials.  Shaded regions indicate one standard deviation around the mean. }
   \Description[]{Correlation with commonality when downsampling the dominant category.}
    \label{fig:red-borda-comm}
\end{figure}

\paragraph{Downsampling all categories}
In our final analysis, we measure  how metrics correlate with commonality when editors downsample items from categories to have similar size. We scale the size of the categories such that 0\

\begin{figure}[t]
\centering

\includegraphics[width=\textwidth]{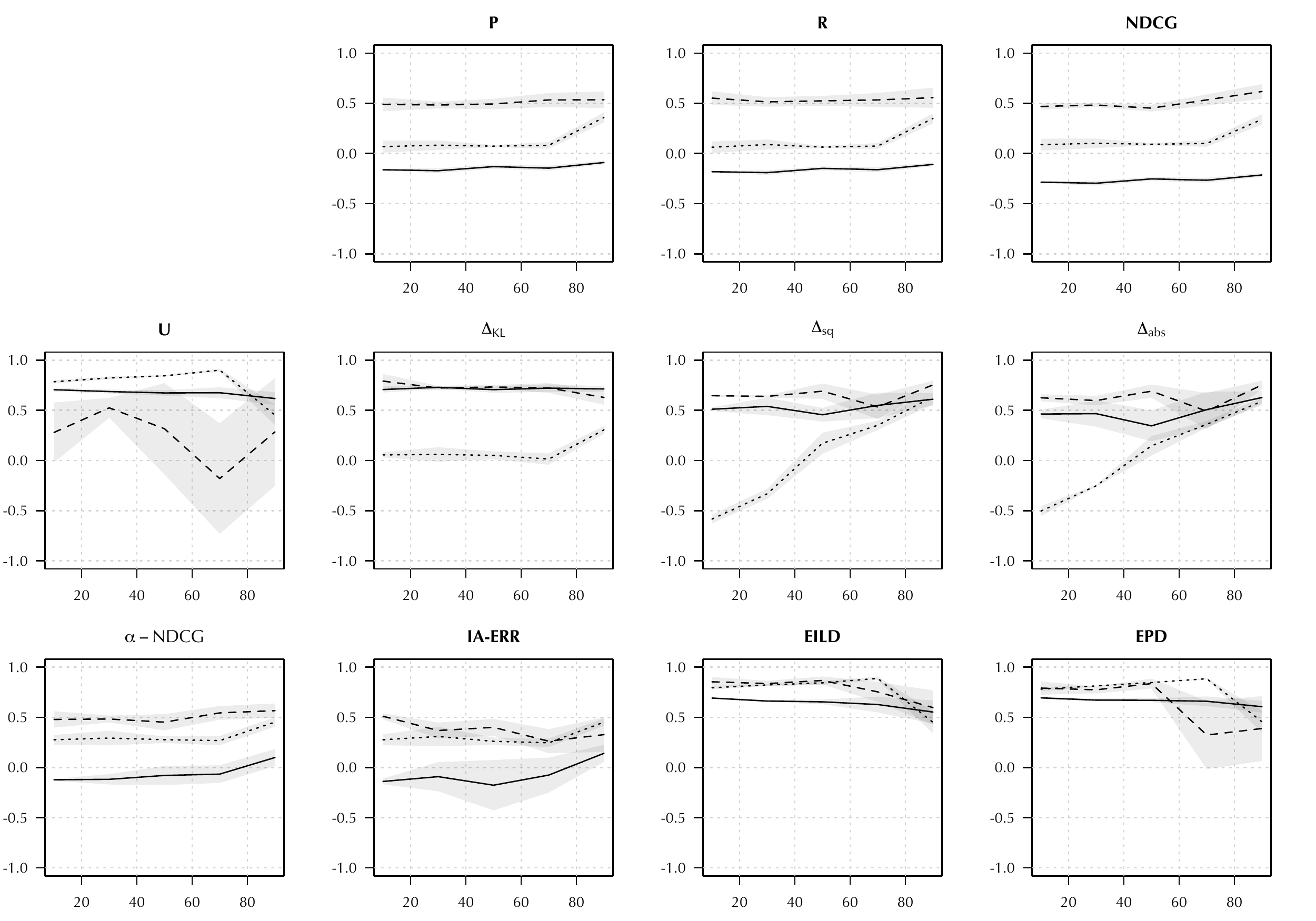}
   \caption{Correlation with commonality when downsampling all categories.  Horizontal axis (all plots): percentage downsampled. Vertical axis (all plots): $\tau$ correlation with commonality. Top row: Utility metrics; middle row: fairness metrics; bottom row: diversity metrics. Solid line: movies; dashed line: music; dotted line: literature. Lines show mean across five trials.  Shaded regions indicate one standard deviation around the mean. }
   \Description[]{Correlation with commonality when downsampling all categories.}
    \label{fig:all-red-borda-comm}
\end{figure}

\paragraph{Summary}
Our analysis indicates that, across the editorial policies we considered,  diversity metrics ($\eild$ and $\epd$) maintain high correlation with commonality.  Although these specific fairness and diversity metrics show stronger correlation than other metrics, their absolute correlation varies substantially across domains and remains relatively far from perfect correlation.  Returning to our earlier analysis, much of this correlation is due to the fact that these measures, unlike utility and fairness metrics, capture the exposure of promoted content \textit{on average} while commonality captures the exposure \textit{simultaneously}.

\subsubsection{Robustness to missing category labels}
\begin{figure*}
  \centering
  \includegraphics[width=\textwidth]{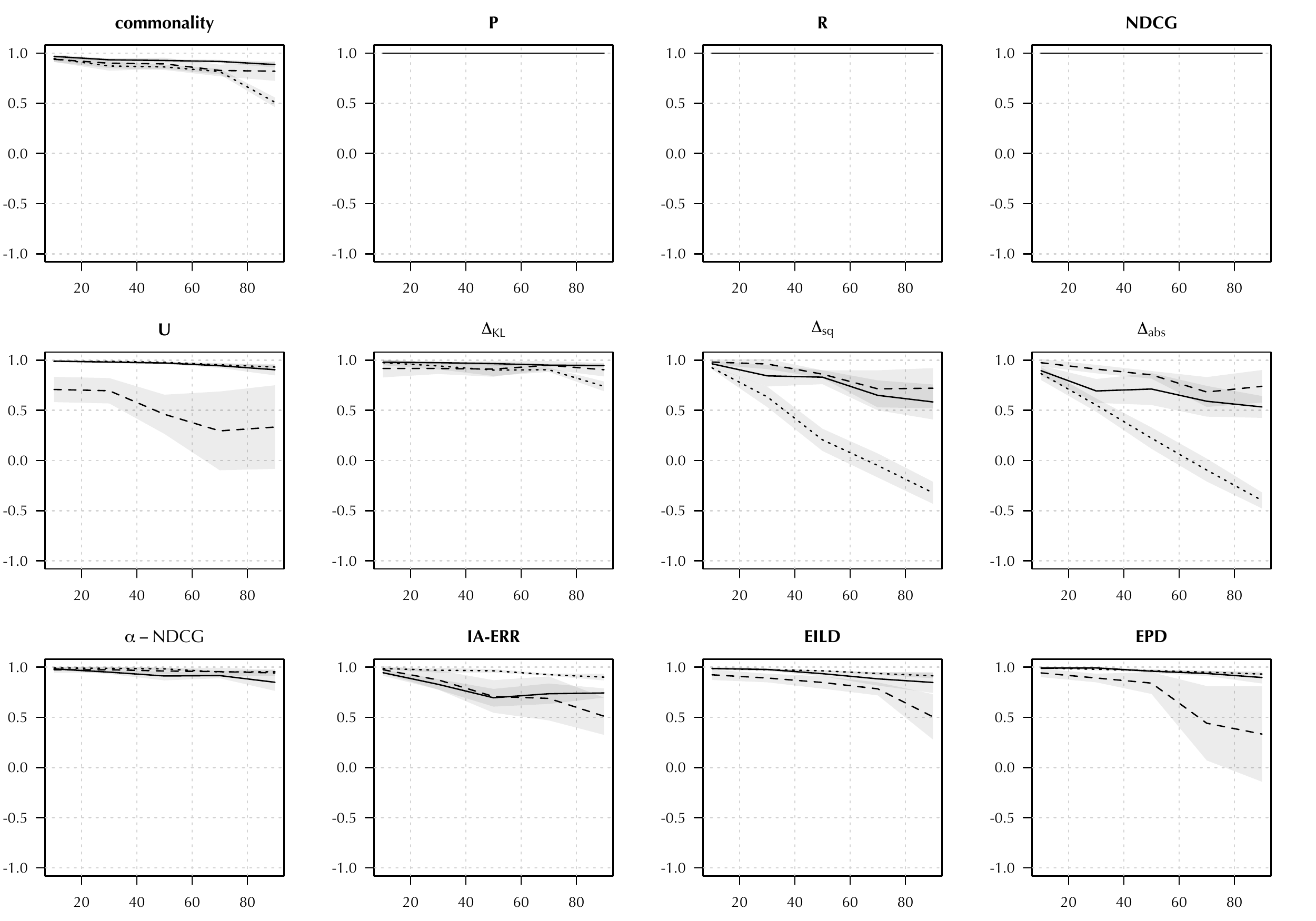}
  \caption{Robustness to missing category labels.  Category labels were progressively removed from items and then the correlation between the system ranking with partial labels and the system ranking with complete labels was measured.    Horizontal axis (all plots): percentage downsampled. Vertical axis (all plots): $\tau$ correlation with complete labels. Top row: Comonality and utility metrics; middle row: fairness metrics; bottom row: diversity metrics. Solid line: movies; dashed line: music; dotted line: literature. Lines show mean across five trials.  Shaded regions indicate one standard deviation around the mean. }\label{fig:labelnoise}
  \end{figure*}

In our second analysis, we evaluate the robustness of commonality  to missing category labels. To do this, we remove category labels for items in each category and measure the correlation between metric computed with incomplete category labels and the metric with complete category labels.  This is different from our earlier analysis because we are simulating errors induced when editors have incomplete information about the complete set of items they \textit{would} select.  We present the results of this analysis in Figure \ref{fig:labelnoise}.  Commonality degrades with increasing label noise due to impact on recall estimates.  That said, since all systems are uniformly subject to incomplete data, the degradation in correlation is slight.  As expected, utility metrics, which do not use category labels show strong correlation with complete label information, regardless of missing labels.  Fairness, diversity, and novelty metrics--with the exception of $\dexpkl$ and $\iaerr$--show more dramatic degradation compared to commonality.

  \begin{figure}[ht]
    \centering
      \includegraphics[width=\textwidth]{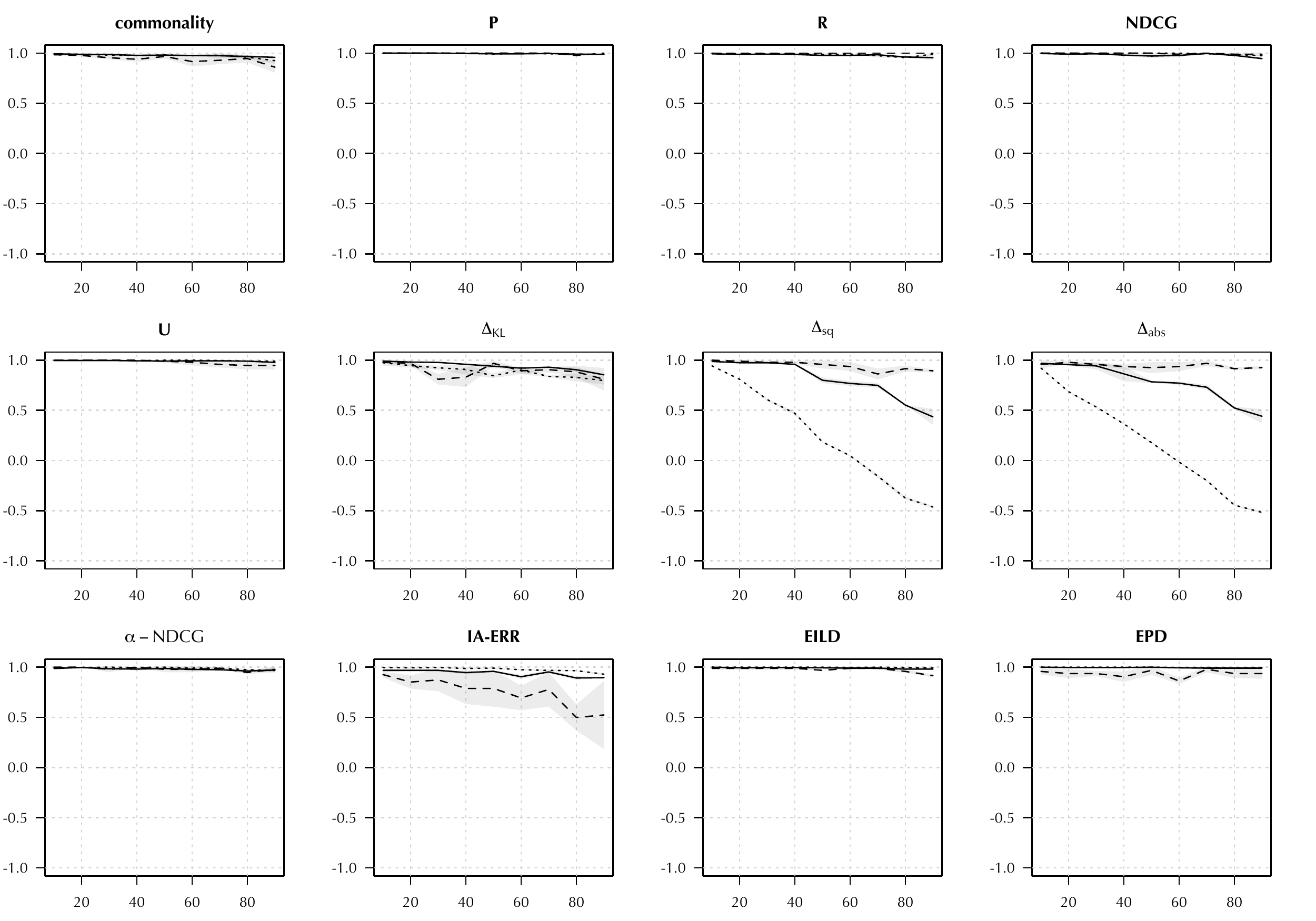}
    \caption{Generalization from sampled users. Correlation between rankings of systems using metrics with samples of users and rankings of systems using metrics with full samples of users.  Horizontal axis (all plots): percentage downsampled. Vertical axis (all plots): $\tau$ correlation with complete labels. Top row: Comonality and utility metrics; middle row: fairness metrics; bottom row: diversity metrics. Solid line: movies; dashed line: music; dotted line: literature. Lines show mean across five trials.  Shaded regions indicate one standard deviation around the mean.  }\label{fig:generalization}
    \end{figure}
  
    \subsubsection{Generalization from sampled users}
    \label{sec:results:missing-users}
Since offline evaluation approximates performance for a full population of users with a sample, we were interested in understanding the stability of commonality under smaller samples. In this analysis, we evaluate commonality on a random subset of users and measure whether the ranking of systems changes significantly. To test how well a metric generalizes to a larger population of users, we randomly sample a subset of users in the range (10-90\

\begin{figure}[ht]
  \centering
  
  \includegraphics[width=0.75\textwidth]{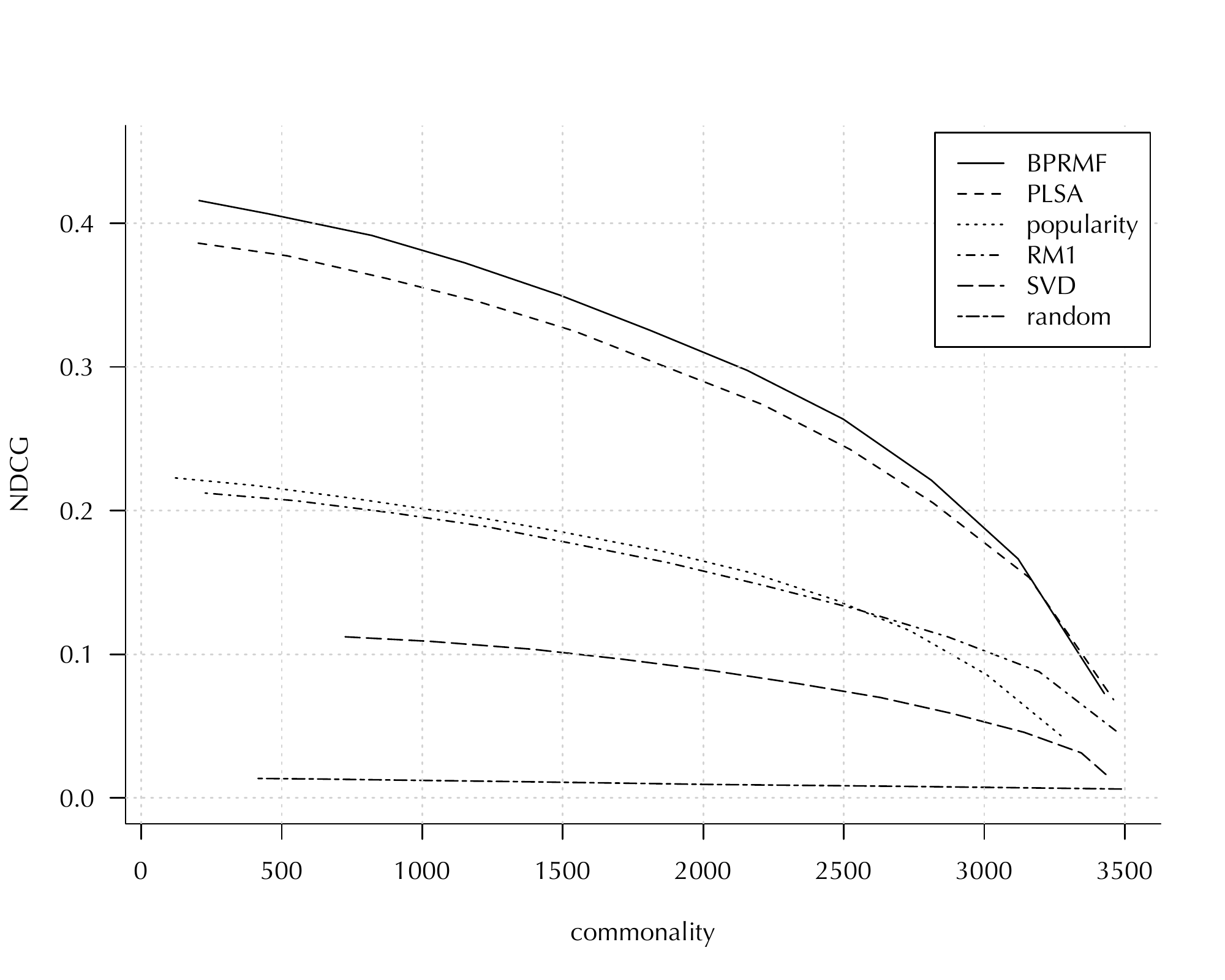}
  \caption{Utility-Commonality Tradeoff Using Interleaved Promotion.  Commonality and NDCG of personalization-focused algorithms post-processed by interleaved promotion.  Results for the movies domain.  }
     \Description[]{Utility-Commonality Tradeoff Using Interleaved Promotion.  Commonality and NDCG of personalization-focused algorithms post-processed by interleaved promotion.  Results for the movies domain.  Curves show a degradation in NDCG as commonality improves.  }
      \label{fig:balance-inter-ncdg-borda-comm}
  \end{figure}

\subsubsection{Improving Commonality of Personalization Algorithms}
\label{sec:metrics:results:mitigation}
So far, we have concentrated on understanding commonality with respect to personalization-focused models.  In this section, we will focus on post-processing the output of these algorithms to make them more commonality-sensitive.  We emphasize simple mitigation strategies so that we can focus on understanding commonality rather on the Pareto-optimal algorithm.  Specifically, we adopt a interleaved promotion algorithm that boosts in-category items within personalization-focused recommendations.  For each user, we order items in a category according to their positions in the personalized ranking and  construct a combined promoted content ranking by selecting items from each category round robin.  We select the top-ranked item in the final interleaved list by sampling an item from either the top of the original personalized ranking with probability $p$ or the top of the combined promoted content ranking with probability $1-p$.  We remove the selected item from its source list.  For second-ranked item in the final interleaved list by sampling an item as before and removing the item from its source list.  We continue this procedure until we have completed the ranking.  A high value of $p$ will recover the original ranking; a low value of $p$ will return the combined promoted content ranking; values in between will be a combination of the two.  We present a detailed description of the algorithm in Appendix \ref{app:balancing}.  We present interleaving results in Figure \ref{fig:balance-inter-ncdg-borda-comm}.  We observe that interleaving allows us to increase the commonality while smoothly degrading utility.  In most cases, tradeoff curves  Pareto dominate each other, indicating that relative utility performance can be largely maintained across commonality targets.  That said, some runs with lower baseline NDCG performance reverse order under interleaving.  This indicates possible systematic under-exposure of content mitigated by interleaving.

\section{Discussion}
\label{sec:discussion}
Since commonality when linked to other progressive cultural principles (here, diversity) is a normative property we seek to promote in recommender systems, we have emphasized clear connections between it and the formal properties of our metric (e.g. diverse curation, familiarity).  This exercise involved substantial translational work between disciplines--between ideas from the social sciences and humanities, and perspectives from recommender engineering. Specifically, we derived normative principles from the literature on public service media and translated them into guiding principles for the design of quantitative evaluation.  In contrast with other evaluation metrics---including many based on personalization--we do not have a latent or delayed quantity to validate the metric.  As such, conceptual analysis and theoretical development play a necessary and an exceptionally important role in the overall research we are presenting here.

We were, in part, motivated by the proposition that existing evaluation metrics fail to capture broader principles associated with the promotion of cultural citizenship.  The mathematical comparison of commonality with existing metrics (Section \ref{sec:metrics:analysis}) demonstrated that formal properties of commonality were absent in existing metrics.  Our empirical results (Section \ref{sec:metrics:results:correlation}) further support this proposition based on the  inconsistent correlation between commonality and existing metrics. Our conviction is that, depending on the cultural context, in principle other normative values could also be formally developed; these, in turn, could be used to assess mathematical or empirical alignment with existing metrics.

Our results in Section \ref{sec:metrics:results:mitigation} demonstrate that existing personalization algorithms can be processed to improve commonality, supporting our proposition that metrics provide an actionable intervention.  While we developed our interleaved promotion algorithm as a proof of concept, more sophisticated algorithms can improve commonality while maintaining high utility.  And, since personalization and public good objectives may be fundamentally in tension, multiobjective methods may be appropriate \cite{mehrotra:balancing}.  

Human editors--and the value communities they channel, and from whom they derive validated categories and judgments--play an important role in the assemblage supporting our evaluation metric.  Even though we used examples of categories justified by existing literature, by envisaging editors answerable to knowledgeable communities that would guide category definition and assignment, we were able to investigate this metric performance while attending in the broader design of the assemblage to SSH concerns about the risks of identity essentialism.\footnote{The risks of essentialism alluded to here are denounced in decolonial data feminist writing, which argues that ``predatory data's algorithmically-driven platforms and `predictive' architectures have massified reductive classification schemes'' \cite{chan:decolonial-feminist-futures}. The alternative envisaged is to promote 'explicitly pluralistic, coalitional knowledge' practices, a version of which we are attempting here through the editors and their relationship to evolving value communities.}   While the categories we selected were limited by labels in our datasets, the general behavior of metrics we observed are representative of the diversity in size and prominence that we  expect in practice.  Given that in this series of experiments, we ourselves substituted for the editors we envisage, we are interested in future work in exploring the extent to which, and how, categories and items selected by those editors would affect the results. Moreover, in some cases, editors may desire more fine-grained control over category importance.  In this situation, we can easily adapt our Borda count method to incorporate weights for categories \cite{ho:weighted-borda}.  

In previous work \cite{ferraro2022}, we aggregated commonality values using mean commonality across groups instead of Borda count (Section \ref{sec:commonality}).  While mean aggregation is appropriate when aggregated values are calibrated across groups, it can degrade in the presence of outliers, which can occur due to differences in category sizes \cite[Figure 1b]{ferraro2022}.  Borda aggregation, on the other hand, preserves only the rank position of each system during aggregation, discarding the magnitude of differences in commonality between runs.  In general, we find that Borda aggregation is necessary to compute a stable aggregation.

\section{Conclusion}
\label{sec:conclusion}

In this work, motivated by defining metrics for recommendation of cultural content, we developed a method to measure alignment with principles of cultural citizenship that we adapted from the PSM literature.
Our proposed commonality metric emphasizes shared familiarity,  the simultaneous exposure of users to content from selected categories.  This definition, captured by the joint probability of familiarity events, is worth exploring for both theoretical, normative, and pragmatic reasons.  

In addition to commonality, we introduced a relatively simple model of familiarity based on recall. We believe there is opportunity to develop alternative models of familiarity that consider a user's previous experience with the category or other contextual information.  However, the design of a familiarity model should be aligned with the concept of shared experience, meaning that, even if a user has engaged with content from a category in the past, \textit{re-exposing} them may promote commonality at the risk of over-satiating users with niche interests, a topic of recent research \cite{leqi:satiation}.  

Our results demonstrate that existing high-utility recommendation algorithms under-perform in terms of commonality.  We believe that exploring the space of commonality-informed recommendation can produce algorithms that perform substantially better in terms of commonality while maintaining high utility.  

By introducing earlier the idea of a recommender system as a sociotechnical assemblage, we point to how future research could attend more to other components of this assemblage, beyond the algorithm, that also bear on diversity or its lack. This might include the catalogues of content on which the system draws, and the larger institutional configuration within which recommender systems are designed and operationalized. Our focus in this paper on the importance of the commonality metric, then, should not be mistaken for the view that developing a new metric is in itself sufficient to advance and achieve the goals we have articulated: recommenders in the public interest that can enhance cultural citizenship.

In future work we are also interested in how the commonality metric attuned to increasing diversity of shared cultural experience might enable us to track these processes over time as, potentially, they cumulatively affect a given population of users. This builds on our founding assumption that existing personalized recommender systems are having cumulative effects--effects that have not yet been identified and studied by the RecSys community. In the same way we assume that the commonality metric could also be tracked over time to identify the cumulative effects of the interventions we describe, making explicit the potential for cumulative changes in cultural exposure among the user population--and potentially bringing to light certain kinds of progressive cultural change.

\appendix
\section{Interleaved Promotion Algorithm}
\label{app:balancing}
\begin{algorithm}[H]
    \caption{Algorithm to incorporate promoted content into a personalization-based ranking.}
    \label{alg:generator2}
    \SetKwProg{interleave}{Function \emph{interleave}}{}{end}
    
    \interleave{User u, Float p}{
         allRecs = getRecommendation ($u$)\;
         List idealRec = new List($items$)\;
         \While {idealRec.size() < 100}{
             List currentCategories = new List($categories$)\;
             \ForAll{Category $c$ in categories}{
                 \If{ $c$ not in currentCategories}{
                     itemC = allRecs.getNextItemByCategory(c)\;
                     idealRec.push(itemC)\;
                     catsItemC = getAllCategories(itemC)\;
                     currentCategories.addAll(catsItemC)\;
                 }
             }
         }
         
        \ForAll{Item $i$ in idealRec}{
            \If{ random number > p:}{
                nextItem = idealRec.pop()\;
            }\Else{
                nexItem =  allRecs.getNextItemNotAdded()\;
            }
            newRec.push($nextItem$)\;
        }
        \Return newRec\;
    }
    \end{algorithm}
    
\bibliographystyle{ACM-Reference-Format}
\bibliography{commonality.bib}
\end{document}